# Death and Maier: meteors and mortuary rites in the eastern Torres Strait

*Eud kerker: na korep maierira asmer opged Torres Straitge*

Carla B. Guedes[1], Duane W. Hamacher[2], John Barsa[3], Elsa Day[3], Ron Day[3], Segar Passi[3], and Alo Tapim[3]

[1] School of Humanities and Languages, University of New South Wales, Sydney, NSW, 2052
[2] Monash Indigenous Studies Centre, Monash University, Clayton, VIC, 3800
[3] Meriam Elder, Mer (Murray Island), QLD, 4875
Correspondence Email: duane.hamacher@gmail.com

**Abstract**

***English***: To the Meriam Mir people of Mer (Murray Island) in the eastern Torres Strait, bright meteors are an important element of death customs and beliefs. We draw from a combination of ethno-historic studies and interviews with Meriam elders to understand the role of bright meteors (Maier) in Torres Strait traditions relating to spiritual elements of death rites using a framework of symbolic anthropology. We find that bright meteors serve as symbolic representations of death and mortuary purification practices and show how the physical properties of meteors are incorporated in ritual, belief, spirituality, and custom.

***Meriam Mir***: Meriamgize maier oditautlare nade eud onagri a mokakalam eud kerker. Kemerkemer daratkapda kikem kerkerira pardar, dorge a oka nako Torres Straitge eud tonar bud ueplare. Debe bibi maieride onatager eud ia onagri a nalu tonar able maierira seri/kakaper a dum able tonar umerem a simir akedrem.

**Introduction**

Natural phenomena often influence the social beliefs, practices, and traditions of cultures in myriad ways (e.g. Call 2012, Chester et al. 2008, Grattan and Torrence 2003), particularly the Aboriginal and Torres Strait Islander cultures of Australia (see Hamacher 2012). Whether literal or symbolic, cultures assign meaning and agency to events such as floods, volcanic eruptions, and earthquakes (Piccardi and Masse 2007). This agency is often symbolic, representing a physical manifestation of ancestor spirits, or denoting divine punishment and malevolent omens (e.g. Hamacher and Norris, 2009, 2010, 2011a, 2011b). The anthropological study of the interaction between humans and their environment, including astronomical and geological events, has a number of practical applications. It can assist in understanding different socio-cultural beliefs and practices with reference to public outreach (Wyatt *et al.* 2014), education (Ruddell *et al.* 2016), Indigenous reconciliation (Tingay 2015), environmental management (Prober et al. 2011), and cultural competence (e.g. Jogia *et al.* 2014).

In the Torres Strait, between Australia's Cape York Peninsula and Papua New Guinea, the Islanders associate some natural events with the symbolic actions of ancestor spirits, which inform social and calendric practices. Early ethnographic work with the Meriam people of the Mer (Murray) Island group in the eastern Torres Strait revealed deep and complex death rites and rituals, connected with ceremony, mummification, and dance (Haddon 1908: 126-162, Hamlyn-Harris 1912, Frazer 1913, Haddon 1935: 322, Aufderheide 2003: 282). Recent interviews with elders and community members on Mer reveal a close relationship between death rites and bright meteors. These relationships and beliefs are very similar to those recorded in the journals of the Haddon expedition in the late 19$^{th}$ century, showing that traditional beliefs about death still survive after 150 years of Christian assimilation.

Using a framework based, in part, on the symbolic anthropology of Geertz (1973) in connection with death and mortuary rituals (Metcalf and Huntington 1991), we argue that bright meteors act as 'vehicles of culture' in terms of both symbolism and physicality. 'Vehicles of culture' are broadly described as devices or means for disseminating or translating culture (Ortner 1984: 129). We find that bright meteors hold cultural meaning and significance that communicates ontological principles related to death, funeral practices, and the transition to the afterlife. We then examine the ways in which meteors inform death rites and act as vehicles of knowledge, noting how the physical properties of meteors inform the observer about a dying person's characteristics, such as their importance in the community, the location of their home, and the size of their family.

**Study Design**

This paper draws from two major sources: ethnohistoric literature (primarily from the Haddon expedition), and interviews conducted on Mer by the second author (Hamacher) from July 2015 to June 2018, as part of Australian Research Council project DE140101600 under the guidelines of Monash University Human Research Ethics project approval code HC15035. All people interviewed were provided a copy of the ethics protocol and signed a participant information consent form. This form specified if the interviews could be recorded via notes, audio, or video and asked permission for interviewees to be named at their discretion. Fourteen Meriam elders and community members were interviewed on multiple occasions (through directed and semi-directed interviews). This involved recording a general conversation about astronomical knowledge, with guided thematic questions posed by Hamacher to call attention to a topic, but not lead questions so as to minimise or eliminate interviewer bias and projection. Elders were interviewed multiple times between 2015 and 2018 during multiple trips to the island. Although participant observation by Hamacher was part of the overall research methodology, little of it related directly to meteors or death rites.

This paper uses a more collaborative approach with elders and Meriam community members. Those interviewed could opt to be named and associated with their knowledge or not. Elders who contributed significantly to the paper in terms of content and review were invited to be co-authors and make contributions at their discretion. Although they were not involved in the direct writing or the use of theoretical frameworks, they did review and approve the original manuscript prior to submission. This is done as a collaborative and decolonising approach to recognising intellectual property and ownership of traditional knowledge, as well as enabling elders to engage in the theoretical interpretation of their knowledge. To highlight the importance of traditional language, the title and abstract were translated into Meriam Mir by Alo Tapim. Future publications will endeavour to include entire manuscripts in both English and the relevant Indigenous language where possible.

Elders and community members who chose to have their identity revealed are (alphabetically) John Barsa, Aaron Bon, Elsa Day, Ron Day, Segar Passi, and Alo Tapim. Within the paper, interviewed elders are referred to by their full name and the year of the interview from which that content was shared with Hamacher.

**The Torres Strait Islanders**

Torres Strait Islanders are an Indigenous Australian people who inhabit the archipelago of islands between Cape York Peninsula and Papua New Guinea. Islanders speak an Aboriginal-based Kala Lagow Ya language and its various dialects in the western and central islands, and the Papuan Meriam Mir language in the eastern islands (Fig. 1), though the two languages share roughly 40% of the same vocabulary. People from the southwest group, including the largely populated islands of Waiben (Thursday Island) and Ngurupal (Horn Island), identify as Kauareg Aboriginal people, but maintain close connections to the broader Torres Strait communities.

The ancestors of the Islanders arrived from southeast Asia over 65,000 years ago when Australia and New Guinea were a single landmass called Sahul (Clarkson et al. 2017). Approximately 8,000 years ago, rising sea levels flooded the land bridge between Australia and New Guinea, forming the western and central islands of the Torres Strait (Straus et al. 2012). The volcanic eastern islands were inhabited approximately 3,000 years BP by Papuan people from the Fly estuary (Carter 2001). Meriam oral traditions speak of four brothers who sailed south from New Guinea (Sharp 2003). A strong wind caused them to separate and the canoe of one of the brothers, Malo, sank at Mer. This brother named the land, its biological features, and established the traditional law, becoming the primary ancestral figure to the people of Mer. This story was recounted and confirmed by several of the elders interviewed.

Torres Strait Islanders maintain close connection and trade networks between the Papuans of New Guinea and the Aboriginal people of mainland Australia (David et al. 2004). European contact with the Islanders began with the arrival of Luis Vaez de Torres of Spain who sailed through the strait in 1606. In 1791, the British HMS Pandora came next, captained by Edwards (Singe 1979). Edwards named the island trio of Mer, Dauar, and Waier the collective term "Murray Islands", which could be considered as a symbolic act of dispossession and abnegation of Meriam sovereignty. Matthew Flinders mapped the region in 1802 on the HMS Investigator. An influx of people from across the Asia-Pacific region came to the Straits on the back of the booming pearl industry in the 19th century, which employed several hundred people by 1877 (Sharp 1993).

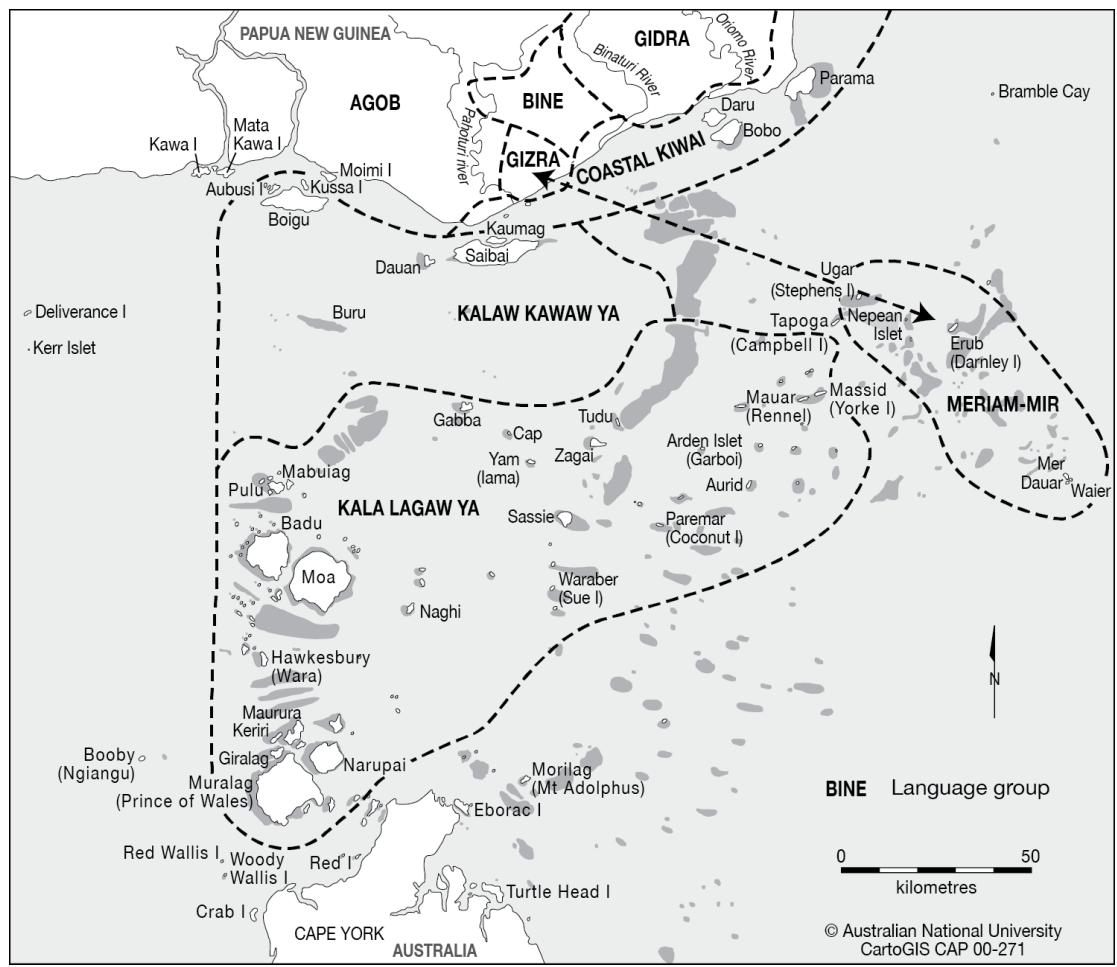

*Figure 1: a linguistic map of the Torres Strait, showing the languages and dialects used in the five major islands groups, and connections to Australia and Papua New Guinea. Image: CartoGIS Services, College of Asia and the Pacific, The Australian National University*

On 1 July 1871, the London Missionary Society arrived on Erub, a date celebrated in the Torres Strait as "The Coming of the Light". This set the stage for Islander assimilation into Christianity, which remains strong across the Straits today. Two major expeditions from Cambridge University reached to the Torres Strait in 1888 and 1898. These expeditions, led by Alfred Cort Haddon, remain the most detailed ethnohistorical records of the Islanders prior to the twentieth century, though the information collected and the methods used are the subject of contention and debate (see Nakata 2007). It is from these journals and archival materials that we find some of the early accounts of Torres Strait Islander traditions linking death rites to meteors. That said, little information on this subject is detailed in these journals and they are obviously written from the perspective of Haddon and the English team rather than the Islanders themselves. It is the goal of this paper to expand upon the role of meteors in Meriam death rites and explore how this phenomenon informs social and ritual beliefs.

**Meteors and Death Rites in Meriam Traditions**

In arguing that meteors play a major role in death rites and beliefs in Meriam traditions, it is important to understand the broader context of Indigenous astronomical knowledge and how this informs social and cultural practices in the Torres Strait. To Torres Strait Islanders, cultural ways of being and ways of knowing are closely linked to the stars (Sharp 1993, Nakata 2010). The Sun, Moon, and stars inform navigation, calendars, weather prediction, seasonal change, food economics, law, ceremony, and social structure (Sharp 1993, Eseli 1998, Hamacher *et al.* 2017). This importance is represented on the Torres Strait Islander flag as a navigational star (Fig. 2).

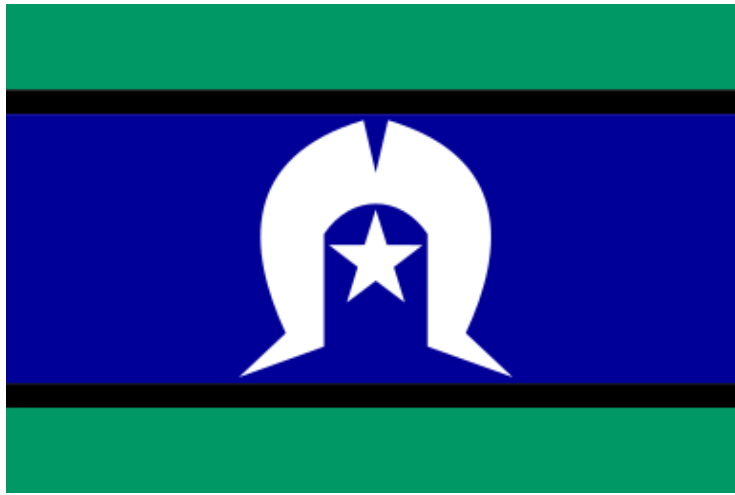

*Figure 2: The Torres Strait Islander flag, designed by Bernard Namok in 1992 and granted Flag of Australia status under the Flags Act 1953 on 14 July 1995.*

The relationship between Islanders and the stars encompasses a sense of identity and belonging – to the natural world, to Islanders' understanding of themselves, and to their cultures. This sense of belonging links the past, present, and future into a holistic system of knowledge that developed over thousands of years (Nakata 2007). This knowledge informs Islander traditions: the laws, customs, and practices that are recorded and handed down in the form of story, song, dance, ceremony, and material culture (Beckett 1987).

Islander astronomical knowledge also contains practical information about the natural world and encompasses how Islanders perceive and understand the environment in which they live (Nakata 2010). In many Australian Indigenous Knowledge Systems, some ritual practices and cultural beliefs are associated with transient celestial phenomena, such as aurorae, cosmic impacts, meteors, eclipses, and comets (Hamacher 2013a, Hamacher and Norris 2009, 2010, 2011a, 2011b, respectively). These associations are typically linked with death, omens, disease, and malevolent beings. In many Aboriginal and Torres Strait Islander cultures, meteors often represent the souls of the recently deceased or evil spirits that hunt for the souls of the living (Hamacher and Norris 2010). As we discuss in this paper, meteors are an important component of this knowledge in Meriam traditions, having relevance to death and mortuary practices, with information about the recently deceased informed by the various physical properties of bright meteors. This paper seeks to address three main questions: (1) What role do meteors have in Meriam death rites? (2) Have these views changed from the days of the Haddon expeditions in the late 19[th] century to today? (3) What do the various properties of stars tell us about traditional views as related by current interviews with elders?

Meriam people call bright meteors *Maier*. They represent the spirits of the dying or recently deceased and possess supernatural powers, such as the ability to cause sickness or death (Haddon 1908: 252-253; all elders interviewed). Early records describing death rites and mourning in Meriam cultures come from the published volumes of the Cambridge Anthropological Expedition to the Torres Strait, led by Alfred Cort Haddon (Haddon 1908: 126-162, Haddon 1935: 322, Hamlyn-Harris 1912). These records are written through the eyes of the non-Indigenous authors and presented in language familiar to 19$^{th}$ century British anthropologists (see Nakata 2007).

Christianity has played a significant role in Meriam society since the arrival missionaries in 1871 (what Islanders refer to as "The Coming of the Light", represented by the colour white in the Torres Strait Islander flag). But many of the pre-Christian traditional views remain to this day. As we shall explore, interviews with elders in relation to death and astronomical phenomena align closely with the accounts described by the Haddon expedition more than a century ago, showing that the influence of Christianity has not extinguished these traditions. Personal beliefs and views may vary from person to person within the community, but all interviewed elders spoke of the same meaning, agency, relationships, and experiences with respect to death and meteors.

Theoretically, descriptions of death and mortuary rites are generally divided between the physical and spiritual remains (e.g. Davis 2008). While these elements work together, the processes for how each are transferred to the next state of being are different. They can include burial, preservation, or decomposition of the physical body and transferal of the spirit to the afterworld. For example, on Saibai in the northwestern Torres Strait, Islanders see a deceased person existing simultaneously in multiple states of being. The person exists in the body form, in the spirit world, and in the commemorations of the community (Davis 2008: 253-254).

Although no longer practiced today, traditionally the physical remains of the deceased Meriam people underwent a detailed, ritual process of mummification and were kept by their families for a period of time (Alo Tapim 2016). Haddon (1908: 253) noted that the Meriam "kept their dead in remembrance as far as their limited resources permitted." This served to temporarily appease the dead person's spirit (Moore 1984) and preserve of the body until the spirit was permanently settled in the afterlife (see Haddon 1935: Plate VII). Moore claimed the mortuary ritual period can last in excess of six months, during which various ceremonies were carried out to finalise the transferal of the soul to its spiritual home. After this point, the mummified body no longer served "religious significance" and was disposed of by the family in various ways.

Haddon dedicates considerable description to death and mortuary practices in the western islands (Vol. V, Haddon 1904) and the eastern islands (Vol. VI, Haddon 1908), focusing largely on mummification, handling of physical remains, and social rules regarding death. The transformation of the person's spirit to the afterworld is addressed to a degree, but a key element – that of the relationship between transcendence of the spirit from the physical world and bright meteors – is almost wholly missing. This study will focus on the spiritual element of death rites on Mer relating to bright meteors.

Ethno-historical documents explain that spirits of the dead are seen by Meriam people as being both playful and something to be feared (Haddon 1908: 253). In the Torres Strait, it is considered relatively common for the living to interact with the spirits of the dead (c.f. Saibai Island; Davis 2008, Sharp 1993: 171). Ancestors sometimes interact and communicate with the living through the Maier. Ron Day (2016) explains:

> *"We went over to Glasgow in Scotland to repatriate ancestral remains. We were led to a building by a group of people. Everyone stopped outside. They said 'go in.' I was a chairman at the time and Sari Tabo was my deputy. We went in there and the room was partially empty except for a few boxes. A man was packing boxes to get them on the road from Glasgow to London. So we went in there and the fellow said to us, look, I'll be out that door so if you need anything, just call me through that door there. So we went and my conscience sort of attracted to that spiritual presence. I looked at Sari and I said to him "can you feel that". The room filled with something like smoke. I think it was the spirits. I said to Sari 'I'll try to contact them through the Maier song'[1]. So I started singing the Maier song. Also, when you're in a traditional situation like that, in Glasgow, you get all the spirits to come home first, then you go home. So I started with that song. And when I finished the room became empty of smoke, it was normal. I said to Sari 'Well, they're gone.' And Sari said yes. It's to do with the falling star. They always imitate – they like to play their games. They use the falling star as a way to signal their presence, say goodbye, or say hello to people. That kind of thing. When I came here, I talked to some people here on the island. When someone dies, when you see a falling star or Maier, you always hear a sound in that direction. It's like they received him and the drums go boom. The drums were beating. I asked someone and that person said 'oh yes, we heard that.' The falling star is part of that."*

When a Meriam person dies, their spirit is taken by the ancestor-being *Terer* to Beig, the land of the dead. Beig is a spiritual place near Boigu in the far northwest of the Torres Strait (Ron Day 2016; Alo Tapim 2016). It is the resting place of the spirits of the dead when going to, or returning from, their traditional home (Haddon 1908: 231). Tapim recounted part of the Terer story, in which he dies and his spirit travels across the different islands of the Torres Strait, finishing at Beig. His mother followed him to bring his spirit back to Mer and revive him.

---

[1] Meriam traditions of the Maier are reflected in song. Elder Alo Tapim sang and translated two Maier songs in the Meriam language. The song has an accompanying dance, which is performed on special occasions on Mer. A Maier (sp. Mayirr) dance performed by Boys and men from Djarragun College was filmed for Mabo Day in the 1990s can be viewed here: www.youtube.com/watch?v=H6nIFs4504w

Haddon records that recently deceased spirits can "haunt" the neighborhood for months after death. Elaborate ceremonies were held during this period for this reason. Haddon groups these funeral ceremonies into two main categories:

1. "A dramatisation of a legend accounting for various practices connected with funerals and the journeying of ghosts to the mythical island of Boigu[2]; in this ceremony the chief performer, who personated *Terer*" (Haddon 1908: 131-132).

2. "The pantomimic representation of recently deceased persons in their character of denizens of the spirit world. We are informed that the illusion of the personification of ghosts by men was almost perfect, more especially as it was assisted by the implicit belief of the women and children that the performers really were ghosts or spirits" (Haddon 1908: 141).

Haddon (1908: 127) reported that spirits of the recently deceased were feared and families would go to great lengths to ensure these traditional ceremonies were conducted frequently and properly so as not to slight the spirits and face their anger or retribution. The ceremonies appear to serve two primary purposes. The first is to reinforce the traditional views of death practices and beliefs through story, song, and action, functioning as a mnemonic. The second relates to mourning practices, serving to comfort the grieving family and community. The ceremony reinforces a timeframe in which families can interact with the deceased, serving as a mode of comfort during a period of grievance. Despite fearing these spirits, the family is content knowing the ceremony will send their spirits home to the afterworld.

Much of this information was addressed and confirmed by Meriam elders during interviews. It is apparent that the views of death and Maier are still passed from generation to generation and largely agree with the basic elements described by Haddon. Contemporary burial practices focus on Christian traditions, but traditional views of the Maier remain. As we will see, ethno-historic literature briefly mentions meteors in connection with death, but fails to go into depth on the subject. From interviews with elders, it is clear that meteors play a more substantial role in Meriam death rites and beliefs, acting as vehicles of knowledge and culture.

**Meteors as Vehicles of Culture**

Ethnohistoric records give some early initial insight to the role meteors play in traditions across the Torres Strait. In the traditions of the Aboriginal Kaurareg people of Muralag (Prince of Wales Island) in the southwestern group, bright meteors are called *titurie udzurra* and are considered *marki* (spirits) that are greatly feared (MacGillivray 1852, Haddon 1890: 434). Sidney Ray confirmed this as a general theme across the Straits (Haddon 1907: 151). The spirits of the dead and their families must be treated with continuous respect, as offending actions can bring "trouble" to relatives. This appeasement can take a general form, as in looking after the deceased's family, or it may take a more specific form, such correctly performing a ceremony or ritual act. If the children of a *Maier* spirit have not been properly looked after by family or the community, the spirit might cast a falling star in the direction of the offending village, resulting in sickness and disease. Sidney Ray noted that while the trajectory of the meteor indicates the location of the place where the sickness or death will occur, time is not a factor - it might happen immediately or it might take some years.

---

[2] Ethnohistoric records and interviews are unclear about the island of the dead. Some say it is Boigu, while others say it is a small island near Boigu called Beig. Haddon (1908: 231) cites Beig as a 'mythological' place underneath Boigu.

The concept of bright meteors being sent as punishment by sky ancestors is a common shared feature in Aboriginal traditions of Australia, and with many Indigenous communities around the world (e.g. Hamacher and Norris 2010). The meteor, usually denoted as being 'fiery', can represent the death (or a portent of death) of an important person, or the physical manifestation of a malevolent spirit or angry ancestor that punishes those who break traditional law. The casting of a *Maier* toward an offending village represents both a physical vehicle of symbolic knowledge and a physical manifestation of the offended spirit.

The act of casting a falling star as punishment is found across Australia and is occasionally used to explain the formation of natural features. Wurundjeri traditions from east of Melbourne describe the sky-deity Bunjil casting down a fiery star as punishment for people breaking sacred law and showing disrespect. The falling star cast by Bunjil created a "bottomless pit" called *Buckertillibe* (Hamacher 2013b). At a place called *Kurra Kurran* at Fennel Bay, Lake Macquarie on the Central Coast of New South Wales, a virtual forest of exposed petrified wood represents the physical remnants of a fiery object cast down by a sky being as punishment for people breaking social taboos in Awabakal traditions (Hamacher and Norris 2009, Hamacher 2013b). Luritja traditions from the Central Desert tell how a fire-devil ran down from the Sun, set the land on fire, created large holes, and killed the people for breaking sacred law. This is the Henbury Meteorite Crater Reserve, a cluster of impact craters that formed a few thousand years ago (Hamacher and Goldsmith 2013; Fig. 3).

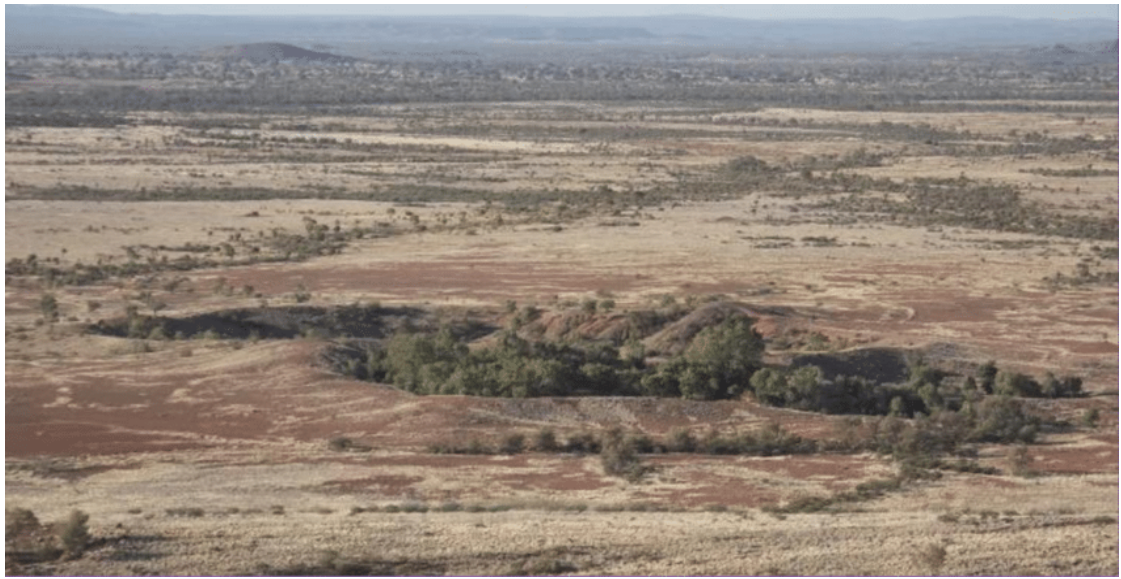

*Figure 3: the larger of the Henbury craters in the Northern Territory. Image: D.W. Hamacher.*

The Henbury impact story represents an oral record of a natural event that would have given impetus to the phenomenon's social meaning. These oral traditions show how Aboriginal and Torres Strait Islander people utilise the landscape to reinforce cultural traditions and enforce sacred law. This utilisation links the sky world, the land of the living, and spirit world and incorporates laws and social customs to a location. This location-specific connection is important to oral cultures, who utilise the *method of loci* to imprint knowledge to memory (e.g. Kelly 2015, 2016).

As explained by all of the Meriam elders and community members who were interviewed, when a person dies or is dying, their spirit is taken to the top of the tallest palm tree by a family member or ancestor (this can be in physical or spirit form). The tree is set alight and the spirit shoots across the sky, traveling to Beig as a *Maier*. Senior Mer Ranger Aaron Bon (2015) referred to this as "soul travel". While on Beig, rituals and ceremonial practices are performed over a period of three days. If someone is sick and dying (but not yet dead), their spirit is treated by a traditional healer. Upon the conclusion of three days, spirit returns home. The spirit can travel to and/or from Beig in the form of a *Maier* – it is not necessarily restricted to one way or the other.

Most elders interviewed recounted a personal experience with a *Maier* and learning of someone passing away. These experiences demonstrate the continued influence and perception of this phenomenon, highlighting the importance it maintains for Meriam people today. Elders described how the *Maier* took an anthropomorphic form. This transformed meteors from being a mundane natural phenomenon to possessing personal meaning and spiritual significance. For Meriam people, bright meteors represent the manifestation of a journey and symbolise a rite of passage. Like many cultures, Meriam people conceptualise death as a transition (Haddon 1908; several elders interviewed 2016-2018). This transition is often seen as a journey to an ultimate destination that may culminate in rebirth, ancestral abode, reunion with nature or divinity, or total oblivion (Metcalf and Huntingdon 1991, Robben 2004). In this context, the personification and anthropomorphism of bright meteors is a symbolic method of embodying the passage of an important person to this destination.

Theoretical frameworks that can be utilised for understanding the relationship between meteors and death are varied. Death rituals that were first delineated by Van Gennep (1909). In this framework, death rituals, like all rites of passage, have a three-part structure: (1) *separation*, (2) *liminality* (referring to a threshold between two states of being), and (3) *reincorporation* (Metcalf and Huntingdon 1991). The spirits of the dead must be separated from their social roles as members in the community of the living and enter an undefined 'in-between' state, finally being reincorporated into a new status at the end of their journey.

Meriam traditions (as recounted largely through interviews), this begins when the spirit is launched from a treetop as a Maier and concludes when the spirit returns home. Some cultures have a specific tie-breaking ritual, which symbolises both the end of mourning for the bereaved, and the end of the journey for the spirit of the deceased. Van Gennep's "end of the journey" ritual for Meriam people involves the ritual of a person's spirit returning home from Beig and finding their place in the spirit world. The physical journey to Beig or back home may be represented by the manifestation of the spirit as a *Maier*, but the spirit remains in another dimension that is generally not seen or experienced by the living.

We consider the framework of Hertz (1960), who argues that death is widely perceived as a process and not as an instantaneous event. In Meriam traditions, meteors serve as a conduit between the physical and spiritual worlds as part of this process. In many Indigenous cultures, as with the Meriam, death is the starting point of a ceremonial process whereby a deceased person becomes a celestial 'ancestor'. The person's death is akin to an initiation into a social after-life and resembles a kind of rebirth. Hertz's study focused on the widespread custom of a second funeral, or more correctly, secondary treatment of the remains. One of the main ideological reasons for a double ceremony has to do with the dual nature of a human being - comprising both body and spirit.

According to Hertz, bodies of the dead are usually associated with ritual pollution and sorrow. The first funeral, in both a literal and symbolic sense, seeks to eliminate these polluting and sorrowful aspects. This coincides with the mummification process and the ceremony about the spirit's journey to Beig. The second funeral is focused on the initiation of the spirit into the realm of the ancestors, where they will continue to serve as a cultural resource, watching over and guiding the living. This coincides with the disposal of the physical remains at the conclusion of the process.

We look to Ortner (2016), who identifies four categories of pollution and impurity in purification rites: (1) physiological processes, (2) violence and related processes, (3) social classifications, and (4) anomalies. Categories (1), (3), and especially (4) are most relevant to the *Maier*. Category (1) is represented, in part, by the mummification process of the body and symbolised by the casting of the spirit by fire across the sky as a *Maier*. Fire is often used for purification and cleansing in the death rites of cultures around the world. This serves as a means of transforming the 'negative' to the 'positive' (Mates 2016; Ortner 2016). Elders interviewed described the *Maier* they witnessed as having an anthropomorphic shape in the form of fire, set alight by the fire of the tree from which the spirit departs. The ascension of the person's spirit as a fiery meteor represents the symbolic purification and cleansing of the spirit, while the mummification process purifies, cleanses, and preserves the physical body for the remainder of the period of liminality.

In Aboriginal traditions of the mainland, the use of fire cast down through a falling star by sky beings represents a destructive force. In the examples previously mentioned (Buckertillibe, Kurra Kurran, and Henbury), the casting of the star by the sky ancestor killed the people and set the land on fire for breaking traditional law. Fire serves as both punishment and a means of erasing the polluting and actions of the offenders, purifying the area. The narrative and associated location or landscape feature highlights the importance of obeying sacred law, reinforcing traditional customs and serving as a stark warning to potential lawbreakers.

Social classification (Category 3) is embedded in the various characteristics of the *Maier* itself (discussed in more detail in the next section). For example, the brightness of a *Maier* is proportional to that person's status in the community or their importance to the observer. Category (4) - 'anomalies' - is the most pertinent to the *Maier*, particularly what Ortner describes as 'certain events of nature'. As a rare natural phenomenon, fireballs (exceptionally bright meteors) only one is observed for every 20 to 200 hours of continuous observation (Richardson and Bedient, 2017). Fireballs are a generally unpredictable and surprising sight that can incite fear and awe in the observer. In addition to possessing several varying visual characteristics, fireballs can produce audible sounds (Spalding et al. 2017) or even explode, producing a sonic boom that can be both heard and felt. The combination of visual, audible, and tactual characteristics give impetus to the significance of the *Maier*.

The period of liminal existence in *Maier* traditions can be thought of in terms of *sacred-space* and *sacred-time*. In The Sacred & The Profane, Eliade (1957) describes *sacred time* as a period of religious or spiritual festival where the participant exits the profane (mundane existence) into a religious or spiritual place and time, involving contact with the divine. Eliade introduces the *Illud tempus* to refer to the 'time-of-origins', the sacred time when the World was created. In the context of *Maier* traditions, *sacred time*, the period of liminality, and the concept of *Illud tempus* are synonymous. This is a period in which ceremonies expel pollution and prepare the spirit for the afterlife, to be reborn clean and healed of any sickness. Aaron Bon (2015) explained that during this time, the sick or dying person was treated by a traditional doctor before travelling back home. Meriam consultants describe this sacred-time as lasting three days. John Barsa (2015) described the spirit's time on Beig as a period of celebration.

Eliade's concept of *sacred-space* is where sacred power or wisdom intrudes upon the profane, and where powerful ancestors are present. The existence of a *sacred-place* breaks the homogeny of space and becomes a 'centre of the world' or the 'highest point in the world'. High regions are significant because they are close to the sky, the ancestors, and concepts of a spiritual world, such as 'heaven'. In this sense, the establishment of a sacred space is necessary to give order to the world and reconnect with the origins of the creation. Therefore, a sacred space enables access to powerful ancestors. The tree from which the *Maier* is launched is the *sacred-place* where the journey begins, as a tree is a high-point and connects the profane world (Earth) to the *sacred-space* of creation (Sky). We emphasise the relevance of this by noting that several elders said the *Maier* spirit was taken to the top of the tallest palm tree to be launched to Beig - the liminal *sacred-space*.

The description of death in relation to *sacred-time* seems to differ when discussing the overall process of a person passing to the afterlife, which can last up to several months as described in the Haddon volumes. In the context of the *Maier*, elders stipulate that this only lasts for three days. The distinction between the two is unclear but seems to refer to a period of celebration rather than long-term ceremony. It is possible the liminal period of three days is an example of syncretism, where the teachings of Christian missionaries were integrated into some of the traditional beliefs and rituals. In the context of Christianity, the liminal period of three days may have been influenced by the crucifixion and resurrection of Jesus. In this context, Jesus absolved humankind for its sins through the brutality of the crucifixion (*separation* through Ortner's Classification 2 – violence), then descending to Hell for three days (liminality), where hellfire cleansed the pollution of sin and constant prayer absolved the impurity of humanity's sins. He was then "reborn" clean and pollutant-free (reincorporation), before ascending to Heaven. A ritual period of three days is found in cultures around the world (Bendann 2007), which are often linked to the three days the Moon "disappears" before re-emerging as a thin crescent, new and cleansed (New Moon). For comparison, many Aboriginal cultures see the Moon as a being who dies for three days before coming back to life as punishment for misdeeds (e.g. Hamacher and Norris 2011).

Although the *Maier* is seen as a physical vehicle symbolising the transition to and from the liminal space or the spirit world, one unnamed elder (2015) recounted an event that coincides with Ortner's description of anomalous behaviour. Years ago, a man was found dazed and incoherent. He claimed to have travelled as a *Maier*, indicating the extreme effects of the meteor (fire, rapid movement) were placed on his corporeal body, not his spirit. According to the account, the man was a skilled dancer who attracted the attention of two spirit women. They fell in love with him but were jealous of his wife. They and tried to take him to the spirit world as a *Maier*, even though he was alive. Some cultures describe the after-effects and dangers of a corporeal body journeying through a medium normally relegated to the spirit-body, explaining the confused and disoriented nature of the man after traveling as a Maier. This was not described by the unnamed elder as a cautionary tale, but rather an experienced event witnessed by the elder and other family members.

**Information Encoded in Meteors**

The *Maier* is perceived as a symbol. It possesses cultural meaning that helps Meriam people overcome the grief of death and the transferral of the deceased person's spirit to the afterworld as a form of personal and social acceptance. But the symbolic meaning of bright meteors extends beyond its mere presence in the sky. The *Maier* is also a physical manifestation of spirit with agency, possessing properties that inform the observer about that person. All of the Meriam people interviewed explained how the various physical properties of *Maier* are tied to cultural belief, linking its physical characteristics to various aspects of the dead or dying person's life. These include the meteor's timing, brightness, colour, fragmentation, sound, and trajectory. These properties work together to assign degrees of significance to the person whose spirit manifests as a *Maier*.

The timing of a *Maier* is only loosely proportional to the passing of an important person. It may happen at the time of death (as noted by the personal experiences of some Meriam elders), or a death may occur sometime after the *Maier* is seen, as a form of postdiction, or *vaticinium ex eventu*. This enables the observer to possess an openness in interpreting the *Maier*. In astrophysical terms, *Maier* are fireballs – a term denoted for bright meteors that have a visual magnitude brighter than $-4$, approximately that of Venus (Richardson and Bedient, 2017). The rate at which meteors of a given brightness appear to an observer are not static in time, but vary depending on annual and diurnal variations (e.g. Hankey, 2017). A fireball of magnitude $-6$ or brighter (Fig 4) will be generally visible to a keen observer in ideal conditions over a period of 200 hours of observation, while a magnitude $-4$ fireball will generally be visible every 20 hours of constant observation (Richardson and Bedient, 2017).

In Meriam traditions, the brightness of the *Maier* is proportional to the importance of that person, either to the community or to the observer. This is not necessarily quantified by magnitude; a person becoming a *Maier* may certainly be of greater importance to a particular family member than the larger community. Given the multiplicity of meaning in *Maier*, this property is not critical. The very brightest *Maier* – blazing fireballs that mimic a second Sun – are likely to signify the passing of a well-respected senior elder. It is worth noting that fainter (non-fireball) meteors are not generally considered *Maier*. They have no special significance in Meriam traditions (John Barsa 2015). In other parts of the Torres Strait, such as Murulag, historical documents show that faint meteors are perceived by the Kaurareg as young novices that have not yet learned how to utilise their full powers and are being trained by the older spirits (Haddon 1908: 253). Faint meteors invoked mockery from the people, who laughed and chastised them. But the appearance of a bright meteor causes a sudden change in mood. The people remain silent until a faint meteor is again seen, bringing back laughter and mockery (*ibid*).

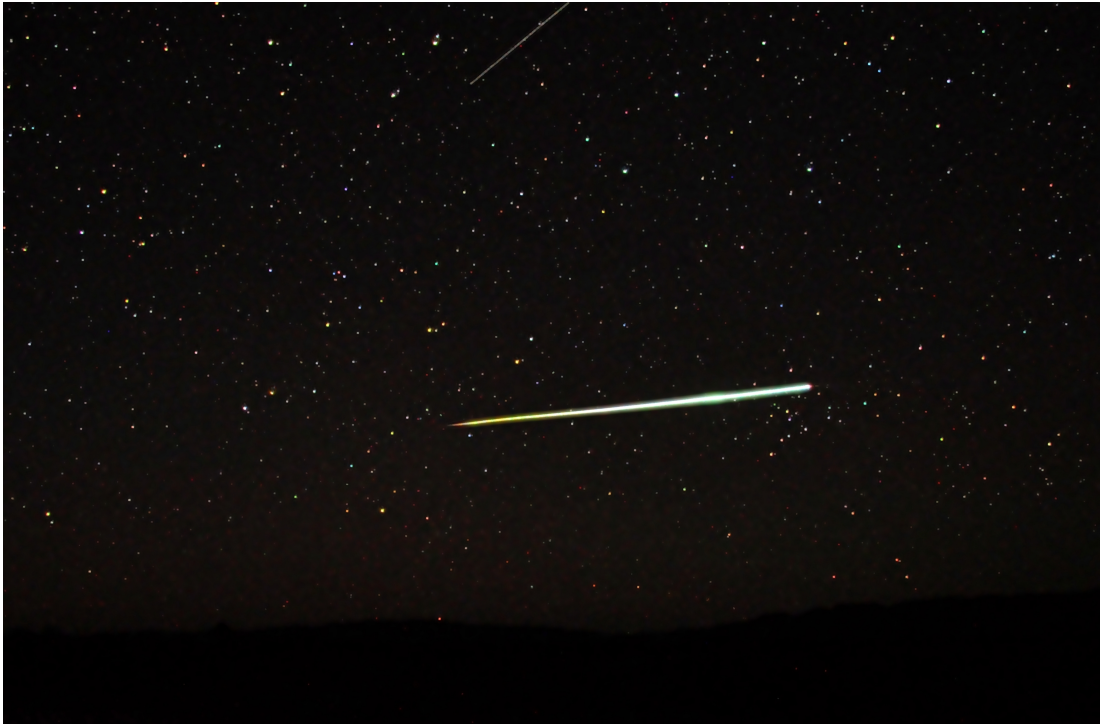

*Figure 4: A bright fireball photographed from the Flinders Ranges, South Australia on 24 April 2011. Image: C.M. Handler, Wikimedia Commons.*

Meteors can appear in a range of colours, including red, white, blue, and green (Opik 2004). In scientific terms, the colour is dependent on the chemical composition of the meteoroid and the temperature at which it burns (Murad and Williams 2002). The variation in colours of meteors seems to have no major role in Torres Strait traditions. Alo Tapim (2015, 2016) explains that the Meriam people associate *Maier* with *ur* (fire). For this reason, *Maier* dances are associated with fire and performed with dancers carrying lit torches (see Hamacher et al. 2017). Haddon (1908: 154) records that the Meriam people also refer to red body paint as maier. Bodies of the dead were painted in whole or part with maier, or it was used to colour motifs carved into the skin. The association with fire is reflected in the placename of a reef between Erub and Mer, called Meiri, or Mary, Reef (divided into a larger and smaller reef). Ron Day (2016) described being able to see flames on or near the reef. He said his parents described it as spirits preparing a Maier and colloquially called it Maier Reef.

If the meteoroid is large enough, generally coinciding with a fireball having a minimum visual magnitude of −8, it may fragment due to increasing ram pressure overcoming the tensile strength of the meteoroid as it transverses the atmosphere (Opik 2004). The fragments falling from the main mass of the meteoroid are described by Alo Tapim (2015) and John Barsa (2015) as 'sparks' in English, or *uir-* *uir* in Meriam Mir. The *uir-uir* indicate the size of the deceased person's family; the more numerous the *uir-uir*, the larger the family. Similar to Meriam traditions, the Kaurareg of Muralag considered a fragmented meteor to mean the spirit had lots of children (Haddon 1890: 434, Haddon 1904: 360, Moore 1979: 156). Fragmenting fireballs are very bright, representing people with large families who are generally more revered and are more likely to have a higher social standing. Thus, a fragmenting fireball reinforces the relationship between brightness and that person's importance in the community.

If a fireball fragments, it is more likely to explode in the atmosphere and cause a sonic boom. In Meriam traditions, this sound is called *dum* - an onomatopoeic term associated with the booming sound of a large drum. All of the elders explained that the *dum* signifies a person's spirit has returned home. Many of the Meriam interviewed described hearing the *dum* when they saw the *Maier*. John Barsa (2015) said he once saw a *Maier* and the accompanying *dum* shake the island, flying towards a creek named Dumapat (meaning *Dum Creek*: *pat* = creek). The *Maier's* trajectory reinforced the relationship between object and place, as well as further supporting the high social standing of that person to the observer and/or community.

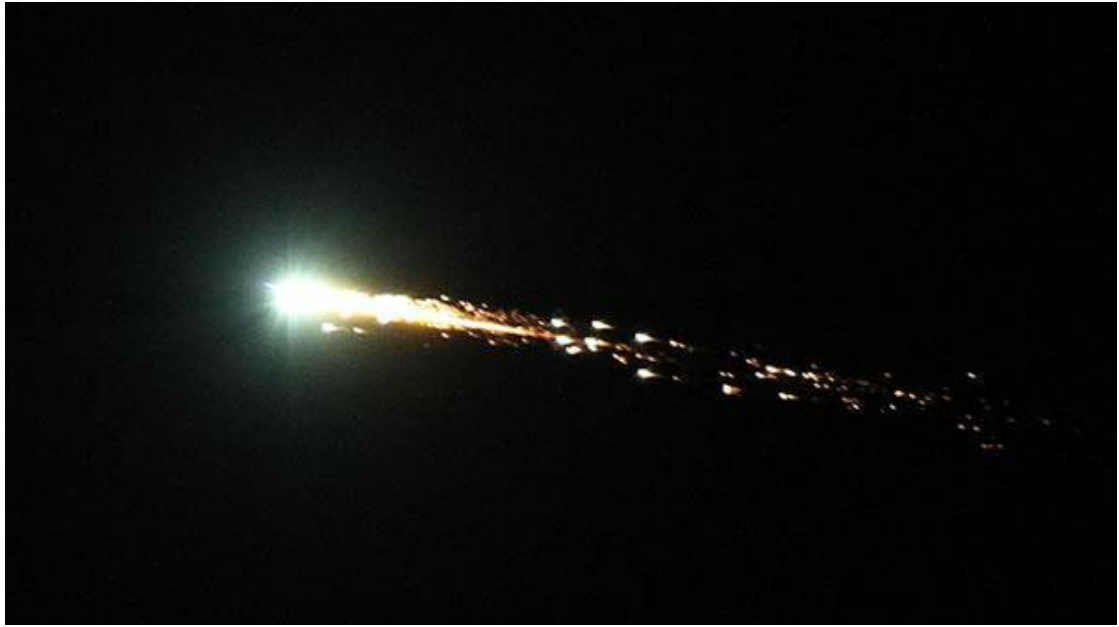

*Figure 5: A fragmenting meteor photographed on 17 October 2012 at Santa Rosa, California, showing what Meriam people call uir-uir, or 'sparks'. Image: NASA/Robert P. Moreno Jr.*

The trajectory of a *Maier* can possess multiple meanings. It may signify the direction to which the spirit is travelling during the period of liminality (e.g. to or from Beig), or it can denote the direction of that person's home. An unnamed elder said the Maier's trajectory indicated the location of the dying person's clan. If the family of a deceased Meriam person are not looked after properly, a *Maier* is cast as punishment by the spirit in the direction of the offenders, thus serving as a directional marker to the offending party's home. A meteor's trajectory can also indicate or reinforce a connection to place. John Barsa (2015) recounted an event in which he and a relative saw a cloud move over the Moon, then a bright Maier with an anthropomorphic shape race across the sky toward Dumapat before exploding, creating the booming *dum* sound. On another part of the island, his brother-in-law claimed to have witnessed a spirit climb to the top of a tree when he was cleaning his catch of fish, followed by the same *Maier* racing toward nearby Dauar island. The next day both men discovered that a Saibai elder had passed away overnight. In addition to demonstrating the significance of a *Maier's* trajectory (and supporting the meaning of creek's placename), this account shows that specific locations in space and time are not necessarily relational to the observed *Maier*. The in-law saw the dying man's spirit climb the tree before appearing as a *Maier*, despite that man being on Saibai in the north of the Torres Strait.

The idea that *Maier* can represent a spirit simultaneously travelling to or from the sacred liminal space, or even possess some other meaning, is a reflection of how Turner (1969: 95) conceptualises liminal entities. He describes them as "neither here nor there." Rather, "they are betwixt and between the positions assigned and arrayed by law, custom, convention, and ceremony." Thus, the meaning assigned to the *Maier* is a reflection of the perceptions of those observing it. This allows for a range of personal interpretations that generally fit within an overall framework, defined by cultural tradition. This can explain how different people seeing the same phenomenon from different locations or angles can perceive and personify it in different ways based on their own interpretations, assumptions, and experiences. Is the spirit travelling to or from Beig? Has the person died or are they dying? Is the trajectory of the Maier telling the observer where the spirit is going, where (s)he is from, or does it represent punishment in the direction of the offender? This concept is reflected in some of the descriptions of the *Maier* as seen by the elders and others at the same time, such as family members, who had similar perceptions of the phenomenon (similar properties, such as brightness, audible sounds, or fragmentation) but different interpretations about its meaning.

**Summary**

We explore the role of bright meteors in Meriam mortuary rites and beliefs, through interviews with elders and archival research. Maier inform observers about the passing of an important person in the community, serving as an important symbol imbedded in rites of passage. Maier represent the ascension of the dead or dying person's spirit to the sky as a form of symbolic purification. They also enable spirits of the dead to communicate with the living, informing people about their life and social status, and even serve as a warning to follow traditional customs. The physical properties of bright meteors - including brightness, colour, trajectory, and sound – inform the observer about various characteristics about the person and their life. Maier serve as symbolic vehicles of culture, and knowledge of their meaning and purpose are passed down through oral tradition and material culture.

The arrival of the London Missionary Society to the Torres Strait, which had a major impact on traditional ways of living, with Christianity influencing views of traditional death rites. But many of the traditional beliefs of the *Maier* remain today, embodied through personal experiences of Meriam people. Concepts and experiences linking *Maier* with death remain in the consciousness and periphery of the Meriam people despite 150 years of ongoing colonisation and religious assimilation.

**Acknowledgements**

We respectfully acknowledge the Meriam elders and participants who shared their knowledge. The interviewed elders and community members on Mer (in alphabetical order) are John Barsa, Aaron Bon, Elsa Day, Ron Day, Segar Passi, Alo Tapim, and four elders who wished to remain anonymous. We dedicate this paper to Uncle John Barsa, who passed away in February 2018, prior to publication of this paper. We thank those who assisted in the project in various ways, including (alphabetically): Kat Henaway, Bob Kaigey, Robert Kaigey, Fleur Kellenbach, Leah Lui-Chivizhe, the Mer Gem Ked Le, Martin Nakata, Aven Noah, Doug Passi, Michael Passi, Tommy Pau, Ken Thaiday, and the Barsa family.

This research was funded by Australian Research Council project DE140101600 under the ethnographic fieldwork guidelines of University of New South Wales and Monash University Human Research Ethics project approval code HC15035. Documentation was obtained from the Haddon Collection at the Cambridge University Library, the British Museum, National Library of Australia, Brisbane Museum, Museum of Northern Queensland, Australian Museum, and Gab Titui on Thursday Island.

This knowledge is part of the heritage of Australian Indigenous peoples, which is fundamental to their spirituality and identity. The two lead authors acknowledge that the anecdotes, knowledge, and songs described in this article are the intellectual property of the elders and knowledge holders. Knowledge holders who contributed to this paper and agreed to be acknowledged are listed as co-authors.